\documentclass[titlepage, 12pt]{article}
\usepackage{graphicx,epsfig, subfigure}
\usepackage{cite}
\usepackage{multicol}
\begin{document}
\begin{titlepage}
\begin{center} 
\textbf{IMAGING CARDIAC DYNAMICS USING LOW-COST ULTRA-HIGH-POWER LIGHT EMITTING DIODES AND VOLTAGE-SENSITIVE DYES} \\
Hana M. Dobrovolny,$^{*1}$ Hany Elmariah,$^{1}$ Soma S. Kalb,$^{2}$ John P. Wikswo, Jr.,$^{3}$ \\ and Daniel J. Gauthier$^{1,2}$ \\
Departments of $^{1}$Physics and $^{2}$Biomedical Engineering, Center for Complex and Nonlinear Systems, Duke University, Durham, NC, 27708 \\
$^{3}$Department of Physics and Astronomy, Vanderbilt Institute for Integrative Biosystems Research and Education, Department of Biomedical Engineering and Department of Molecular Physiology and Biophysics, Vanderbilt University, Nashville, TN, 37235
\end{center}

\textbf{ABSTRACT:} We describe the characteristics of low-cost ultra-high-power light emitting diodes (LEDs) for use in optical imaging experiments. We use the LEDs in experiments with bullfrog cardiac tissue and find that the signal-to-noise ratio is comparable to other commonly used illumination sources.\\

\begin{flushleft} 

\underline{Contact Information:}\\

Duke University, Department of Physics, Box 90305, Durham, NC, 27708 \\
%%\vspace{-13pt}
Tel: (919) 660-2512 \hspace{.1in} Fax: (919) 660-2525 \hspace{.1in} email: hdobrovo@phy.duke.edu \\
\end{flushleft}
\end{titlepage}

\LARGE
\noindent
Introduction
\normalsize

The development of voltage-sensitive dyes has revolutionized the study of electrical activity in spatially extended biological systems such as the heart \cite{Rosenbaum:01}. In an optical mapping study, the electrical activity at different spatial locations can be visualized directly using a fluorescent dye in combination with an illumination source to excite the dye and a detector array to record fluoresence. A typical excitation source for use with di-4-ANEPPS (specifications for other dyes are available from Molecular Probes, www.probes.com) needs to provide an intensity of $\sim$10-100 mW/cm$^{2}$ at the tissue surface, a spectral bandwidth less than $\sim$35 nm, and the variation in the power of the source over time must be much less than the anticipated change in fluorescent power when the cell depolarizes.  

Some groups have recently investigated the use of light emitting diodes (LEDs) as illumination sources \cite{Kodama:00, Entcheva:02}. The narrow bandwidth, high efficiency, and the potential for low-noise operation of LEDs satisfy the illumination source requirements for successful optical imaging. However, these early experiments used low-power (0.25-10 mW) LEDs so they could only image small areas of cardiac tissue. 

In this paper, we report on the use of recently available ultra-high-power LEDs (Lumileds, model Luxeon Star/O and Star/V) as an illumination source in cardiac optical mapping systems. Our study is needed because the LEDs operate at high power where thermal effects can be important and noise in the higher current supplies can be difficult to control. Specifically, it is not obvious that they will operate with sufficiently low noise to be useful in biological imaging experiments. 

The ultra-high-power LEDs are available in two different models: Star/0, lower power (35-85 mW) with collimating optics, and Star/V, higher power (200-400 mW) without collimating optics. The new LEDs are significantly more powerful than those used in previous experiments and are available in a variety of emission wavelengths \cite{Luxeon:star} that span the near-infrared, visible, and near ultraviolet spectrum, so the LEDs can be used with a variety of different dyes and for ratiometry experiments \cite{Tsien:90}. These LEDs are also significantly less expensive ($\sim$\$15-\$40) than either lasers or white light sources. Thus, they offer an attractive option for use in optical imaging experiments if they can achieve a signal-to-noise ratio (SNR) comparable to currently used sources.

\LARGE
\noindent
Methods
\large

\noindent
\emph{LED characteristics}
\normalsize

We measured the intensity of the LEDs with a New Focus photodetector (model \#2031) at various distances for both the Star/O and Star/V models. Both models were run at their maximum rated current (700 mA for the Star/V and 350 mA for the Star/O) by a low-noise, constant current power supply (Agilent model E3615A).

The LEDs must not only provide enough intensity for the experiments, the intensity must remain stable over the course of the experiment. Since these LEDs are so much more powerful than previous models, heating of the junction may affect light output. We mounted the LEDs on CPU heat sinks and fans to help dissipate heat. We then turned on the LEDs and recorded the intensity with the New Focus detector for 20 seconds every 2 minutes over a span of 10 minutes, and every 10 minutes thereafter for an hour. 

Finally, we measured the noise of the LEDs by illuminating a card stained with fluorescent paint that mimics the fluorescence of the dye in an optical mapping experiment, but without the constant disruption of action potentials. We used a DALSA (model CA-D1-0128T) camera to collect 500 images at a frame rate of 490 Hz. We calculated the mean and standard deviation of the intensity for each pixel to determine the relationship between intensity and noise.

\large
\noindent
\emph{In vitro Experiments}
\normalsize

We used the LEDs to perform optical mapping experiments in pieces of bullfrog ventricular myocardium. These studies were performed in accordance with the Research Animal Use Guidelines of the American Heart Association and the Public Health Service Policy on Humane Care and Use of Laboratory Animals, and the experimental protocol was approved by the Duke University Institutional Animal Care and Use Committee. Two bullfrogs were anesthetized and their hearts were excised. Small pieces (about 5$\times$5$\times$3 mm) of ventricular myocardium were stained with 50 $\mu$M di-4-ANEPPS and placed in a tissue chamber. The tissue was superfused with standard Ringer's solution and paced at a constant basic cycle length. In each of the experiments, we illuminated the tissue with either a cyan or green LED. 

Since both LEDs emit some light at wavelengths greater than the cut-off for our high-pass filter, additional dichroic filters (Edmund Industrial Optics H52-538 and H52-535) were placed in front of the LEDs to block any long wavelength emission. We used an OG 590 filter (Edmund Industrial Optics H46-064) to filter the fluorescent emission and we captured the images with the DALSA camera. We made several 3,000-frame recordings to determine the signal amplitude and signal-to-noise ratio (SNR) of both sources. 
%%\pagebreak

\LARGE
\noindent
Results
\large

\noindent
\emph{LED Characteristics}
\normalsize

\begin{figure}
\begin{multicols}{2}
\centerline{\textsf{A}}
\epsfysize=2.5in
\centerline{\epsffile{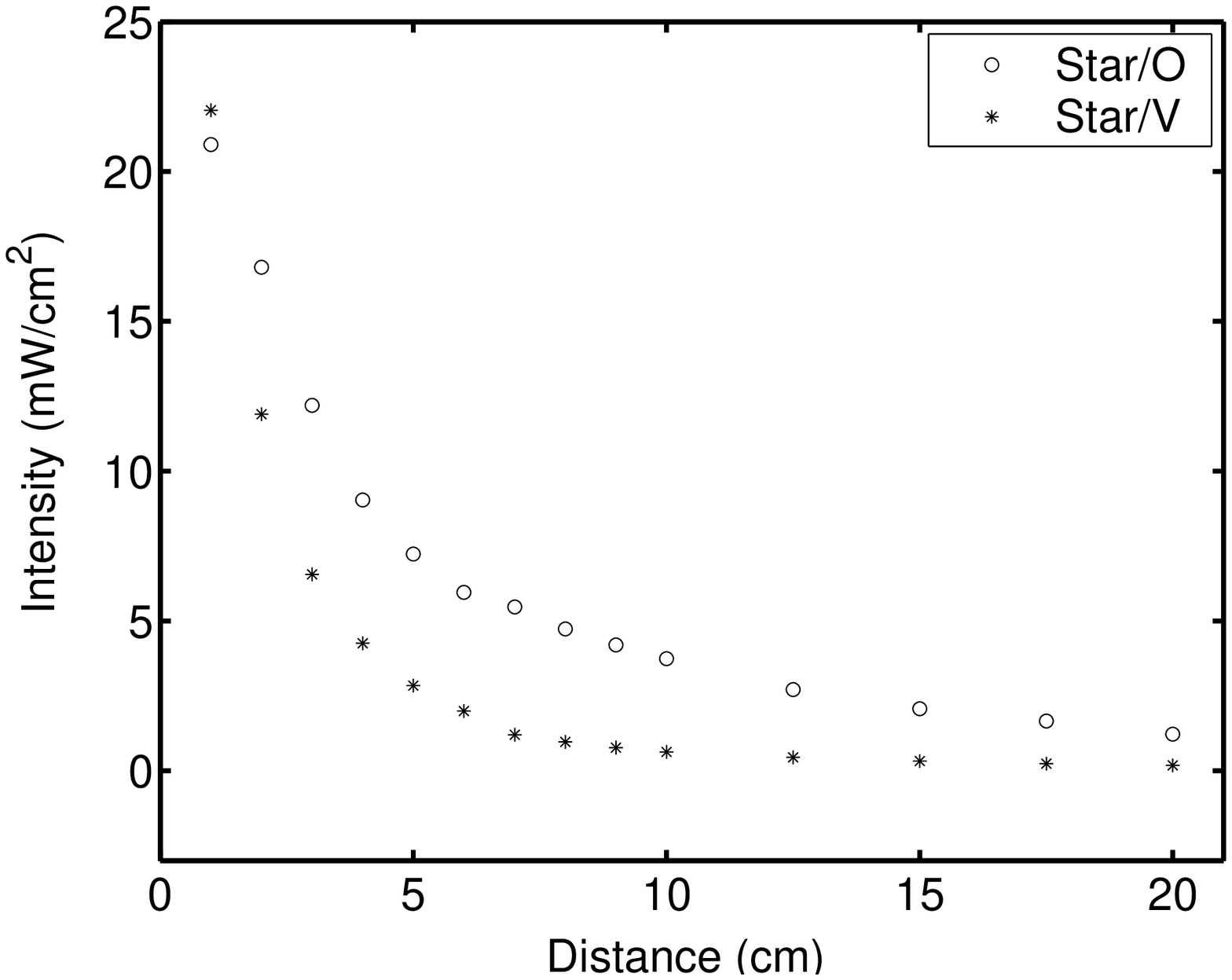}}
\centerline{\textsf{B}}
\epsfysize=2.5in
\centerline{\epsffile{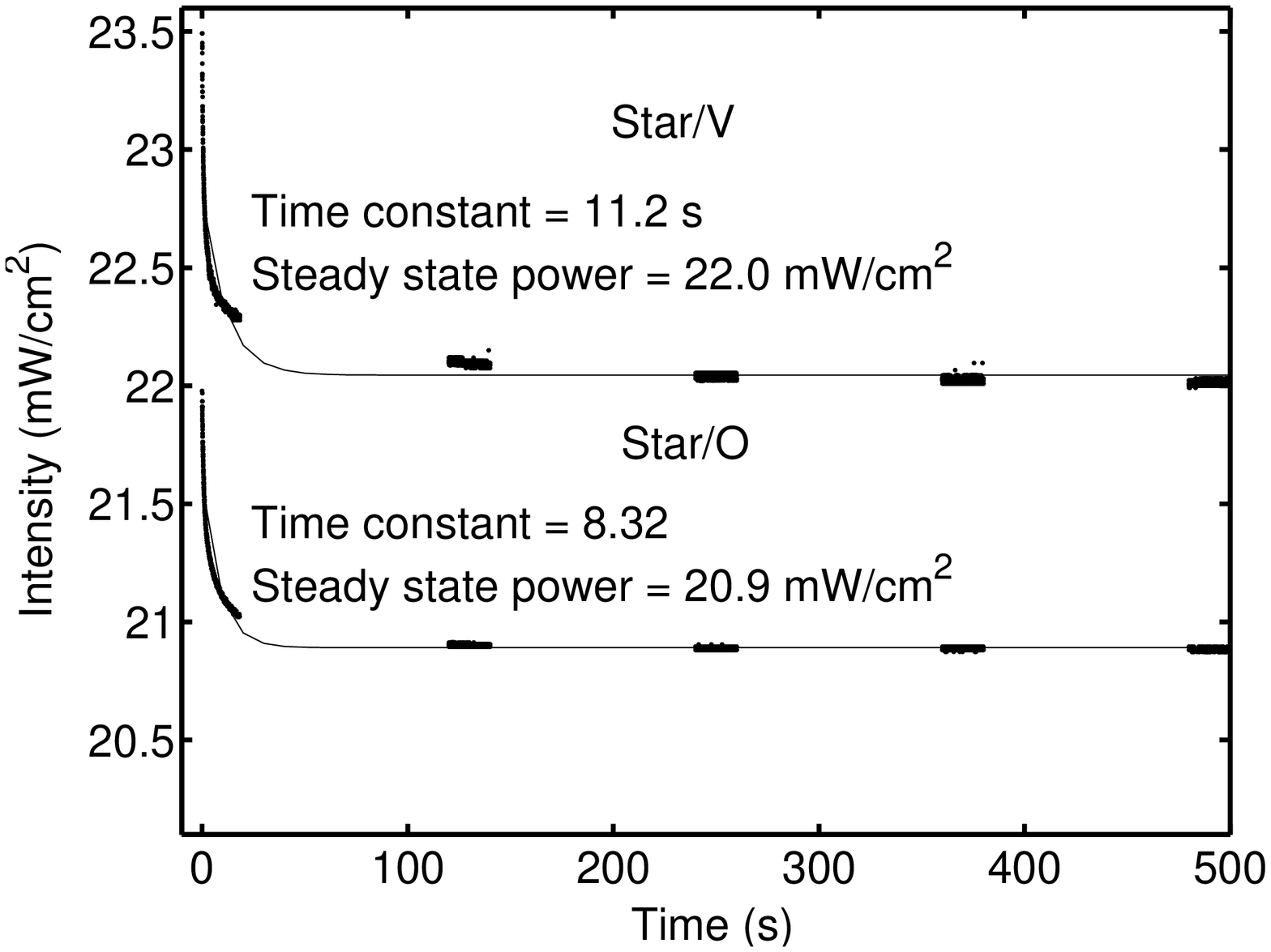}}
\end{multicols}
\caption{\label{distance}}
\end{figure}

Figure \ref{distance}A shows the intensity as a function of distance of the green LED (results are similar for other colors). The Star/O manages to produce intensities greater than those of the Star/V (more powerful) LED once we are more than $\sim$1 cm from the source. Thus, for imaging experiments where the source needs to be some distance from the tissue, the Star/O LED is the better choice.\footnote{We have recently learned that the plastic lens that is an integral part of the Star/O devices can be purchased separately from Luxeon. The lens can be glued to the Star/V devices, leading to much higher intensities in comparison to the Star/O devices.}   

A time course of the intensity after initial turn-on (Figure \ref{distance}B) shows that the intensity output of the LEDs is quite stable. We see an initial drop in light output of 6.3\% for the Star/V and 5.0\% for the Star/O. 

Although this is a fairly large drop in intensity (a drop in fluorescent intensity close to an action potential), it only happens over a short period of time after the initial turn-on; the time constant for the Star/V is 11.2 s and for the Star/O it is 8.3 s. The LEDs reach steady state within 30 s and the light output remains very stable (within the digitization error of our data acquisition card: $\pm$0.001 mW/cm$^{2}$) beyond the initial decrease in intensity. 

Finally, measurements from the fluorescent card show that the LEDs operate near the shot noise limit. 

\begin{figure}
\begin{multicols}{2}
\centerline{\textsf{A}}
\epsfxsize=2.5in
\centerline{\epsffile{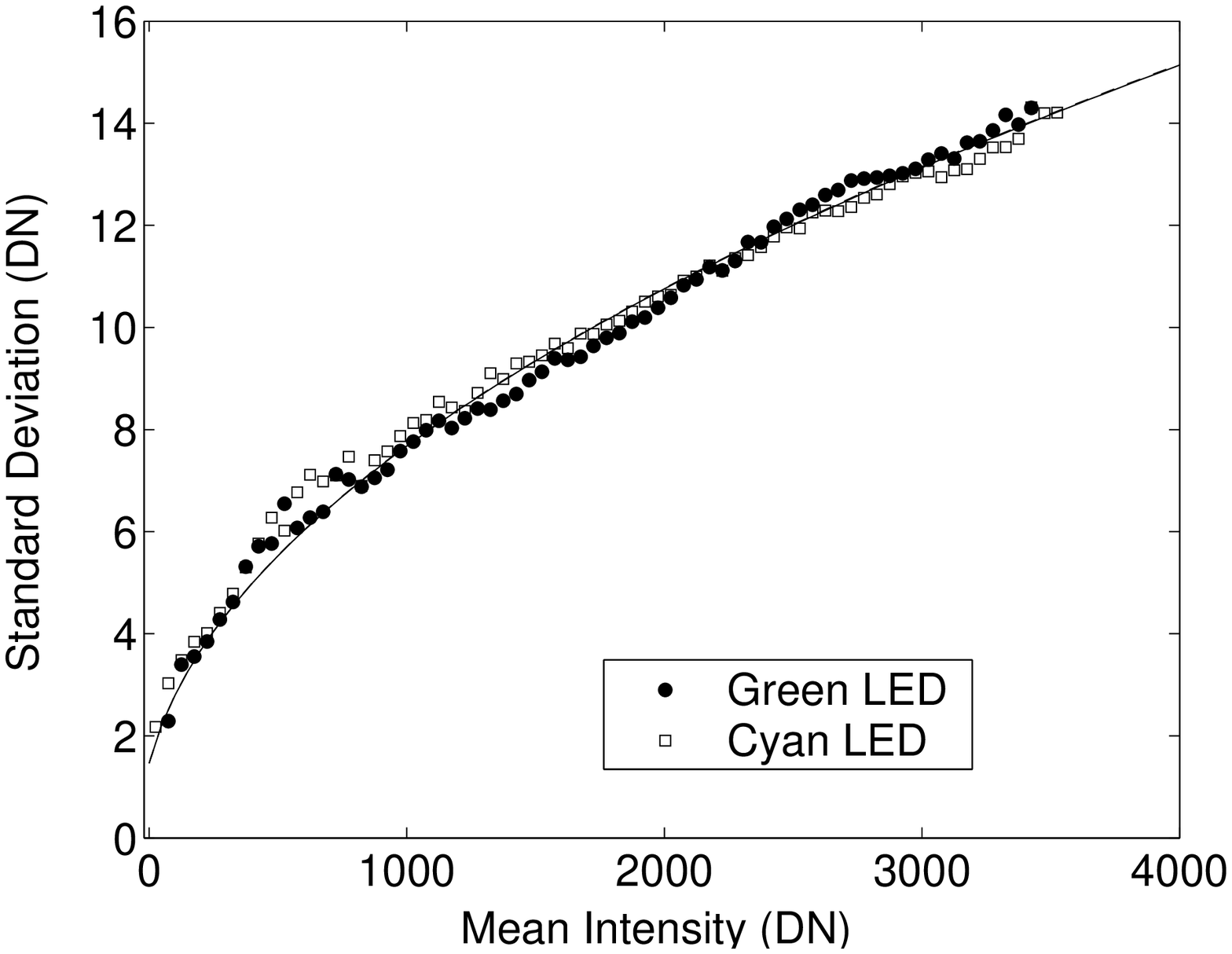}}
\centerline{\textsf{B}}
\epsfxsize=2.5in
\centerline{\epsffile{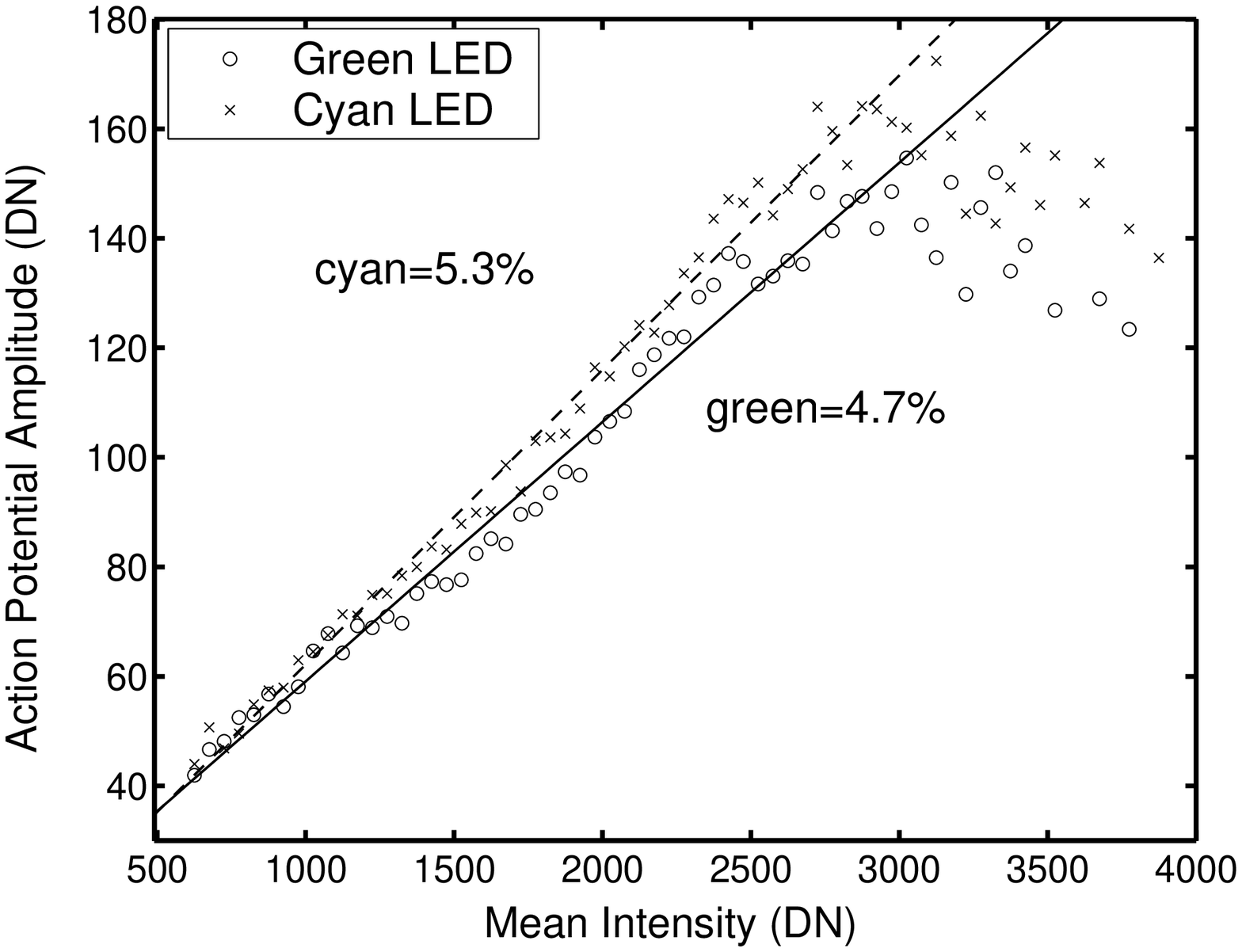}}
\end{multicols}
\caption{\label{amp}}
\end{figure}

We expect an ideal light source to have the following relationship between noise and mean intensity \cite{Boyd:83}:
\suppressfloats
\begin{equation}
\label{theory}
\Delta_{N}=\sqrt{R\cdot\overline{N}+\Delta_{dark}^{2}},
\end{equation} 
where $R$ is the camera conversion factor, $\overline{N}$ is the mean intensity (in digital numbers) and $\Delta_{dark}$ is the dark noise of the camera. Any deviation from the form of this equation indicates the presence of technical noise, such as variation in the light output of the emission source or noise from the power supply \cite{Maolinbay:00}. To determine how well the data fit the above relationship, we fit the mean and standard deviation values for each source to Eq. \ref{theory} (Figure \ref{amp}). Experimentally determined values of $R$ are given in Table \ref{results}. Both the green and cyan LEDs fit the theoretical curve well, indicating that the LEDs operate near the shot noise limit.

\large
\noindent
\emph{In Vitro Experiments}
\normalsize

\begin{figure}
\begin{multicols}{2}
\centerline{\textsf{A}}
\epsfxsize=2.5in
\centerline{\epsffile{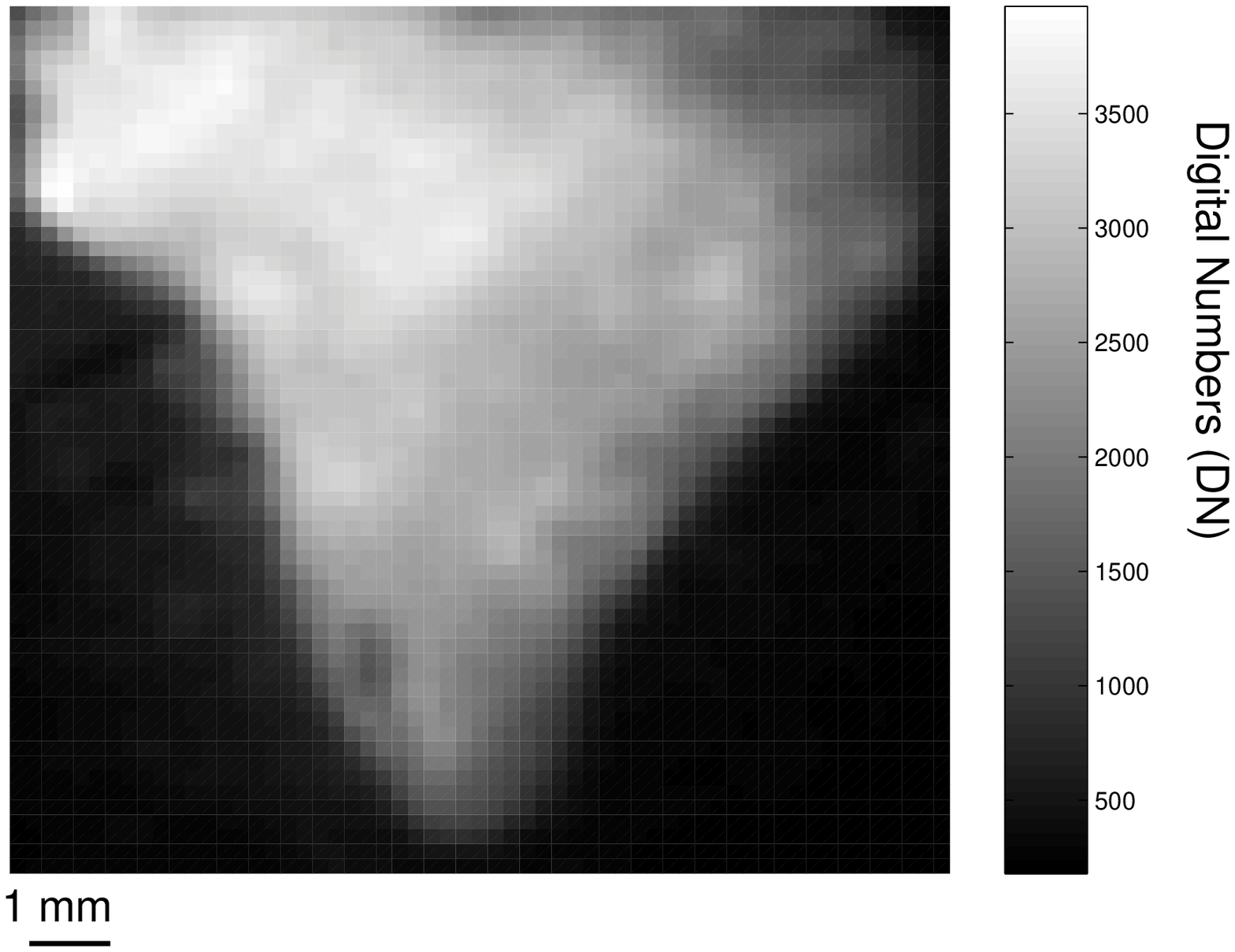}}
\centerline{\textsf{B}}
\epsfxsize=2.5in
\centerline{\epsffile{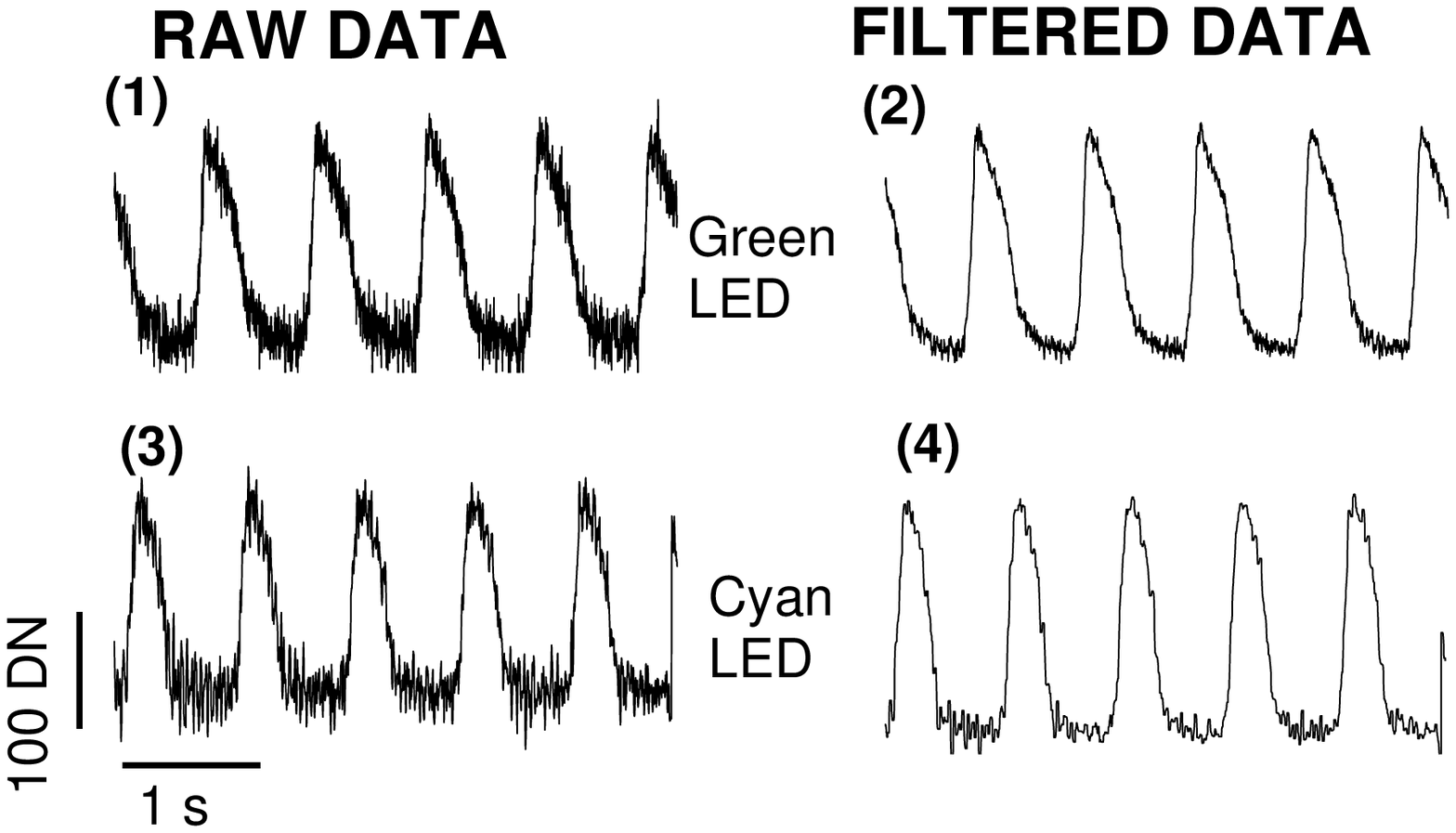}}
\end{multicols}
\caption{\label{aps}}
\end{figure}

Figure \ref{aps} shows representative action potentials from the \emph{in vitro} experiments. Both raw data and filtered data are shown. We determined the average action potential amplitude (APA) for each pixel and results were binned and plotted as a function of mean intensity. We then fit the data to a straight line, the slope of which gives the percent change in intensity during the action potential. Figure \ref{amp} shows the results for both cyan and green LEDs. We find that both cyan and green LEDs perform equally well with similar maximum signal to noise ratios (see Table \ref{results}).

\begin{table}
\centerline{\begin{tabular}{|c|c|c|c|}\hline
Source& R&APA&SNR\\ \hline
Green LED&0.0568$\pm$0.006&4.7$\pm$0.8\%&13$\pm$3\\ \hline
Cyan LED&0.0569$\pm$0.006&5.3$\pm$0.8\%&14$\pm$3\\ \hline
\end{tabular}}
\caption{\label{results} Results of the noise and action potential amplitude (APA) measurements.}
\end{table}

\LARGE
\noindent
Discussion and Conclusion
\normalsize

We have shown that the characteristics of the Luxeon Star LEDs satisfy the requirements for an illumination source in an epifluorescence measurement of transmembrane potential. Both the Star/O and the Star/V provide sufficient stable light intensity for imaging experiments. The SNR in these experiments is comparable to the SNR found in other imaging experiments using lasers or white light sources \cite{Salama:88}. Our experiments easily imaged pieces of tissue of area $\sim$0.5 cm$^{2}$ using a single LED. Use of multiple LEDs permits imaging of even larger areas. Since the LEDs are less expensive, more compact and more efficient than current light sources, evidence of their comparable performance in experiments makes them an attractive new option for optical imaging. 

\LARGE
\noindent
Acknowledgements
\normalsize

This research was supported by National Institutes of Health grants 1 R01 HL072831-01 and 2 R01 HL58241-07, and National Science Foundation grant PHY-0243584.
\nopagebreak
\bibliographystyle{jbiomed}
\bibliography{bibliography}
\pagebreak

Figure \ref{distance}: Intensity of the green LED as a function of (A) distance and (B) time. 

Figure \ref{aps}: (A) Intensity map of a piece of frog cardiac tissue. (B) Optically recorded action potentials from frog hearts. Pacing interval was 800 ms. Raw data (1,3) was filtered with a 3$\times$3 spatial Gaussian filter and three-point temporal averaging (2,4).

Figure \ref{amp}: Signal and noise of the LEDs. (A) Standard deviation as a function of the mean intensity for both cyan and green LEDs as measured from the fluorescent card. (B) Recorded action potential signal as a function of mean  intensity for frog hearts. The slope of the line gives the percent change in intensity during the action potential. 

\end{document}